\documentclass[twocolumn,showpacs,preprintnumbers,amsmath,amssymb,prl]{revtex4}

\usepackage{graphicx}
\usepackage{dcolumn}
\usepackage{bm}

\usepackage{mathptmx, courier, pifont}
\usepackage[scaled=0.92]{helvet}
\usepackage[T1]{fontenc}
\usepackage{textcomp}
\usepackage{color}
\usepackage{colordvi}

\begin{document}


\title{Urca Cooling in Neutron Star Crusts and Oceans: Effects of Nuclear Excitations} 
\thanks{Intended for unlimited release under LA-UR-21-21037.}%

\author{Long-Jun Wang}
\email{longjun@swu.edu.cn}
\altaffiliation{School of Physical Science and Technology, Southwest University, Chongqing 400715, China} 
\author{Liang Tan}%
\affiliation{School of Physical Science and Technology, Southwest University, Chongqing 400715, China}%
\author{Zhipan Li}%
\affiliation{School of Physical Science and Technology, Southwest University, Chongqing 400715, China}%

\author{G. Wendell Misch}
\email{wendell@lanl.gov}
\affiliation{Theoretical Division, Los Alamos National Laboratory, Los Alamos, NM, 87545, USA}%

\author{Yang Sun}
\email{sunyang@sjtu.edu.cn}
\affiliation{School of Physics and Astronomy, Shanghai Jiao Tong University, Shanghai 200240, China}%

\date{\today}

\begin{abstract}
  The excited-state structure of atomic nuclei can modify nuclear processes in stellar environments. In this work, we study the influence of nuclear excitations on Urca cooling (repeated back-and-forth $\beta$ decay and electron capture in a pair of nuclear isotopes) in the crust and ocean of neutron stars. We provide for the first time an expression for Urca process neutrino luminosity which accounts for excited states of both members of an Urca pair. We use our new formula with state-of-the-art nuclear structure inputs to compute neutrino luminosities of candidate Urca cooling pairs.  Our nuclear inputs consist of the latest experimental data supplemented with calculations using the projected shell model. We show that, in contrast with previous results that only consider the ground states of both nuclei in the pair, our calculated neutrino luminosities for different Urca pairs vary sensitively with the environment temperature and can be radically different from those obtained in the one-transition approximation.
  
\end{abstract}

\maketitle

Stars end their lives in different ways leaving different objects as remnants: white dwarfs, black holes, or neutron stars (NSs) \cite{Heger_RMP_2002, Iliadis_book}. Studies on phenomena that are related to stars and these remnants could help us to understand many open questions such as the evolution of the Universe, the origin of elements etc. Neutrinos play crucial roles in these phenomena by various mechanisms \cite{langanke_2020_EC_review}. For example, many astrophysical environments could be cooled effectively by the nuclear Urca process \cite{Gamow_1941_PR}.

\begin{figure}[htbp]  \includegraphics[width=0.42\textwidth]{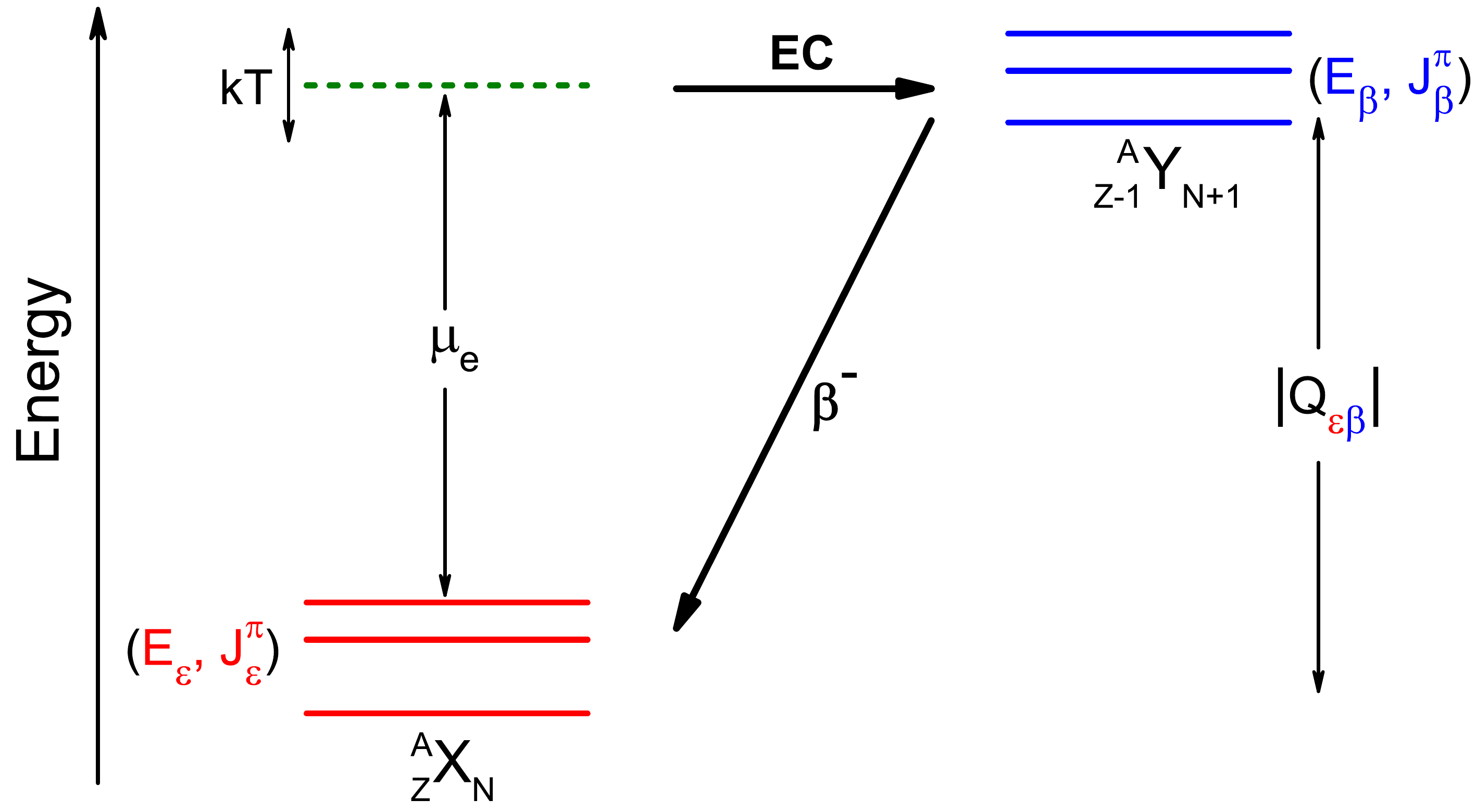}
  \caption{(Color online) Schematic energy diagram for the Urca process. Electron capture (EC) and $\beta^-$ decay cycle rapidly between nuclear states with energies $E \lesssim kT$ when the electron chemical potential $\mu_e$ is sufficient to overcome the $Q$-value ($\mu_e \approx \left|Q_{\varepsilon\beta}\right|$). }
\label{fig:one}
\end{figure}

In the nuclear Urca process, a nucleus releases energy and cools its environment by absorbing an electron and emitting a neutrino (electron capture, EC), then emitting an electron and an antineutrino ($\beta^-$ decay). A large amount of energy can be carried away by the neutrinos produced by these two reactions.
\begin{subequations} \label{eq.Urca}
\begin{eqnarray}
  \text{EC}: \quad ^A_Z\text{X} + e^- &\rightarrow& ^{\quad A}_{Z-1}\text{Y} + \nu_e, \\
  \beta^-: \qquad \  ^{\quad A}_{Z-1}\text{Y}   &\rightarrow& ^A_Z\text{X} + e^- +\bar\nu_e
\end{eqnarray}
\end{subequations}
These reactions occur back and forth between pairs of nuclei (Fig. \ref{fig:one}). It is important in pre-supernova stars (cooling the core leads to greater electron degeneracy and a stronger post-bounce shock in the resulting supernova) \cite{patton2017presupernova}, and it has been considered in the context of white dwarfs \cite{Tsuruta_1970_ASS, Schwab_2017_MNRAS_Urca} and type Ia supernovae \cite{Paczynski_1972_ApJL, Woosley_1986_AR, Stein_2006_ApJ, Toki_2013_PRC_Urca, Schwab_2017_ApJ}.  However, early NS crust models adopted a zero-temperature approximation which precludes the electron phase space necessary for Urca cycling.

Recently, \citet{Schatz_2014_Nature} showed that finite temperatures in the crusts of accreting NSs can open sufficient phase space for electrons to be emitted by $\beta^-$ decay. They found that the Urca process is an effective coolant within a shell of a few meters thickness in the crust, and they identified fourteen dominant Urca pairs. A subsequent study of the NS ocean found fifteen dominant Urca pairs \cite{Deibel_2016_ApJ}. By using analytic expressions based on Ref. \cite{Tsuruta_1970_ASS} with available experimental data supplemented by theoretical values, both studies identified dominant pairs by calculating the total neutrino luminosity in their EC/$\beta^-$-decay cycles. It was found that nearly all Urca pairs consist of odd-$A$ nuclei, because the odd-even staggering of nuclear masses disfavors Urca cycling in most even-$A$ nuclei \cite{Ong_thesis, Meisel_2015_PRL, Meisel_2017_ApJ, Meisel_2018_JpG}.

The above studies made important advances in understanding the Urca process in NSs, but the calculations include transitions almost exclusively between ground states; in some cases, an extremely low-lying state is used as a proxy for the ground state. However, in environments with moderate temperature, nuclei may have considerable probability to be thermally excited, while cold environments imply that even low-lying states will be sparsely populated. Nuclear excitations are expected to strongly modify nuclear reaction and decay rates, with sensitive dependence on density and temperature \cite{Tan_2020_PLB}. In this spirit, \citet{Fuller_1994_ApJ} studied the Urca process with thermal nuclear excitations in supernova core collapse, but there is no comparable NS study. In NS crusts and oceans with typical temperatures $T \lesssim 10^9$ K, excited states with $E_x \lesssim 100$ keV in both parent and daughter nuclei can affect the corresponding Urca cooling.  This possibility demands investigation because odd-$A$ nuclei can have many low-lying states, especially when they are deformed \cite{Schatz_2014_Nature}.

Urca cooling of accreting NSs is usually considered in a full nuclear reaction network \cite{Schatz_2014_Nature, Lau_ApJ_2018}. Although nuclear excitations can substantially contribute to the reaction rates (and hence the neutrino luminosities) in the network, it is complicated and impractical to integrate the phase space exactly when nuclear excitations are taken into account \cite{Fuller_1994_ApJ}. In the present work, we follow Refs. \cite{Tsuruta_1970_ASS, Deibel_2016_ApJ} and derive a modified analytic expression for Urca process neutrino luminosity which includes excited states. With the inclusion of detailed nuclear structure and transitions involving excited states, we find considerable differences compared to earlier results \cite{Schatz_2014_Nature, Deibel_2016_ApJ, Meisel_2017_ApJ}; this may radically change conclusions as to which Urca pairs are the most effective NS coolants. 


From Refs. \cite{Tsuruta_1970_ASS, Schatz_2014_Nature, Deibel_2016_ApJ}, the neutrino luminosity $L_\nu$ of the Urca process can be approximated analytically as
\begin{eqnarray}
  L_{\nu}(Z_{\varepsilon}, A_\varepsilon, T) \approx
  L_{34} \times 10^{34} \mathrm{erg} \ \mathrm{s}^{-1} X(A_\varepsilon) T_{9}^{5} 
  \left(\frac{g_{14}}{2}\right)^{-1} R_{10}^{2}, \label{got_L_nu}
\end{eqnarray}
where $X(A_\varepsilon)$ is the mass fraction, $T_9$ is the temperature in GK, $g_{14}$ is the NS surface gravity in units of $10^{14}$ cm s$^{-2}$, and $R_{10}$ is the NS radius in units of 10 km. The intrinsic cooling strength $L_{34}$ quantifies the effects of nuclear properties:
\begin{equation} \label{previous_L_34}
  L_{34}(Z, A) = 0.87 \left(\frac{10^{6} \ \mathrm{s}}{ ft } \right) \left( \frac{56}{A} \right)
  \left( \frac{ |Q_{\text{EC}}| }{ 4 \ \mathrm{MeV} } \right)^{5} \left(\frac{\langle F\rangle^{*}}{0.5}\right) .
\end{equation}
Here, $Z$ and $A$ are the charge and mass numbers of the EC nucleus, $ft$ is the comparative half-life for the nuclear weak transitions, and $Q_{\text{EC}}$ denotes the difference in \emph{atomic mass} between the nuclei.  The quantity $\langle F \rangle^\ast$ is effectively an averaged Fermi function that accounts for the Coulomb corrections to the electron wavefunctions.
\begin{align}
  \langle F \rangle^\ast &\equiv \frac{ \langle F\rangle^{+} \langle F\rangle^{-} }{ \langle F\rangle^{+} + \langle F\rangle^{-} } 
  \quad \\
  \quad
  \langle F\rangle^{\pm} &\approx \frac{ 2 \pi \alpha Z }{ \left|1-e^{(\mp 2 \pi \alpha Z)}\right| } \label{eq:F_plus_minus}
\end{align}
The superscript $+$ ($-$) denotes EC ($\beta^-$ decay), and $\alpha \approx 1/137$ is the fine-structure constant.

Eq. (\ref{previous_L_34}) considers only one transition (ground state to ground state in most cases), neglecting the thermal population of nuclear excited states in the finite-temperature environment.  To remedy this, we followed Fuller, Fowler, and Newmann (FFN) \cite{Fuller_1980_ApJS, Fuller2, Fuller3, Fuller4} and assumed a Boltzmann distribution for the occupation probability of excited states in parent nuclei to derive a modified version of Eq. (\ref{previous_L_34}):
\begin{eqnarray} \label{got_L_34}
  && L_{34}(Z, A, T) =  \\
  && \sum_{\varepsilon \beta} 0.87 \left(\frac{10^{6} \ \mathrm{s}}{\langle ft \rangle_{\varepsilon \beta}}\right) \left(\frac{56}{A}\right)
  \left[ \frac{ |Q_{\varepsilon\beta}(Z,A)| }{ 4 \ \mathrm{MeV} } \right]^{5} \left(\frac{\langle
  F\rangle^{*}}{0.5}\right). \nonumber
\end{eqnarray}
The index $\varepsilon$ ($\beta$) labels low-lying states of the EC ($\beta^-$-decay) parent nucleus, and $Q_{\varepsilon\beta}$ is the difference in total nuclear energy between the two states $\varepsilon$ and $\beta$ (more on this below).  The effective thermal $ft$ value for individual transitions is
\begin{eqnarray} \label{L_34_ft_if}
  \langle ft\rangle_{\varepsilon \beta} \approx \frac{ \widetilde{f t}_{\varepsilon \beta}^{-} + 
  \widetilde{f t}_{\varepsilon \beta}^{+}  }{ 2 },
\end{eqnarray}
with
\begin{subequations} \label{def_tilde}
\begin{eqnarray}
  \widetilde{ft}^+_{\varepsilon \beta} &\equiv& \frac{G^+(T)}{(2J_\varepsilon + 1) e^{-E_\varepsilon / (kT)}} ft^+_{\varepsilon \beta} , \\
  \widetilde{ft}^-_{\varepsilon \beta} &\equiv& \frac{G^-(T)}{(2J_\beta + 1) e^{-E_\beta / (kT)}} ft^-_{\varepsilon \beta}.
\end{eqnarray}
\end{subequations}
The quantity $G^+$ ($G^-$) is the partition function for the EC ($\beta^-$-decay) parent nucleus:
\begin{eqnarray}
    G^{\pm}(T) = \sum_{\varepsilon/\beta} (2J_{\varepsilon/\beta}+1) e^{-E_{\varepsilon/\beta} / (kT)} , 
\end{eqnarray}
where $E_{\varepsilon/\beta}$ ($J_{\varepsilon/\beta}$) is the excitation energy (spin) of the nuclear level; again, $\varepsilon$ and $\beta$ index the EC and $\beta^-$-decay parent states.  Because the occupation of excited states changes with temperature, the expression in Eq. (\ref{got_L_34}) is \emph{temperature-dependent}, while the previous formulation in Eq. (\ref{previous_L_34}) is not.

Our expression (\ref{got_L_34}) includes two corrections to Eq. (\ref{previous_L_34}). First, by summing over low-lying states, the effects of thermal excitations in both nuclei are included. Second, the $Q$-value should be related to the difference in \emph{nuclear energy} rather than \emph{atomic mass}. That is,
\begin{eqnarray} \label{Q-Q_EC}
  Q_{\varepsilon\beta} &= M_\varepsilon c^2 - M_\beta c^2 + E_\varepsilon - E_\beta \nonumber \\
                        &= Q_{\text{EC}} - m_e c^2 + E_\varepsilon - E_\beta,
\end{eqnarray}
where $M_\varepsilon$ ($M_\beta$) is the nuclear mass of EC ($\beta$) parent nucleus. This implies a difference of $m_e c^2$ with respect to $Q_{EC}$. For Urca pairs in the NS crust and ocean, $Q_{\varepsilon\beta}$ and $Q_{\text{EC}}$ are negative. Considering the case of only ground-state-to-ground-state transition as in Eq. (\ref{previous_L_34}), we have $|Q|=|Q_{\text{EC}}|+m_e c^2$. This correction generally increases the neutrino luminosity even without considering excited states. 

\begin{figure}[htbp]
\begin{center}
  \includegraphics[width=0.45\textwidth]{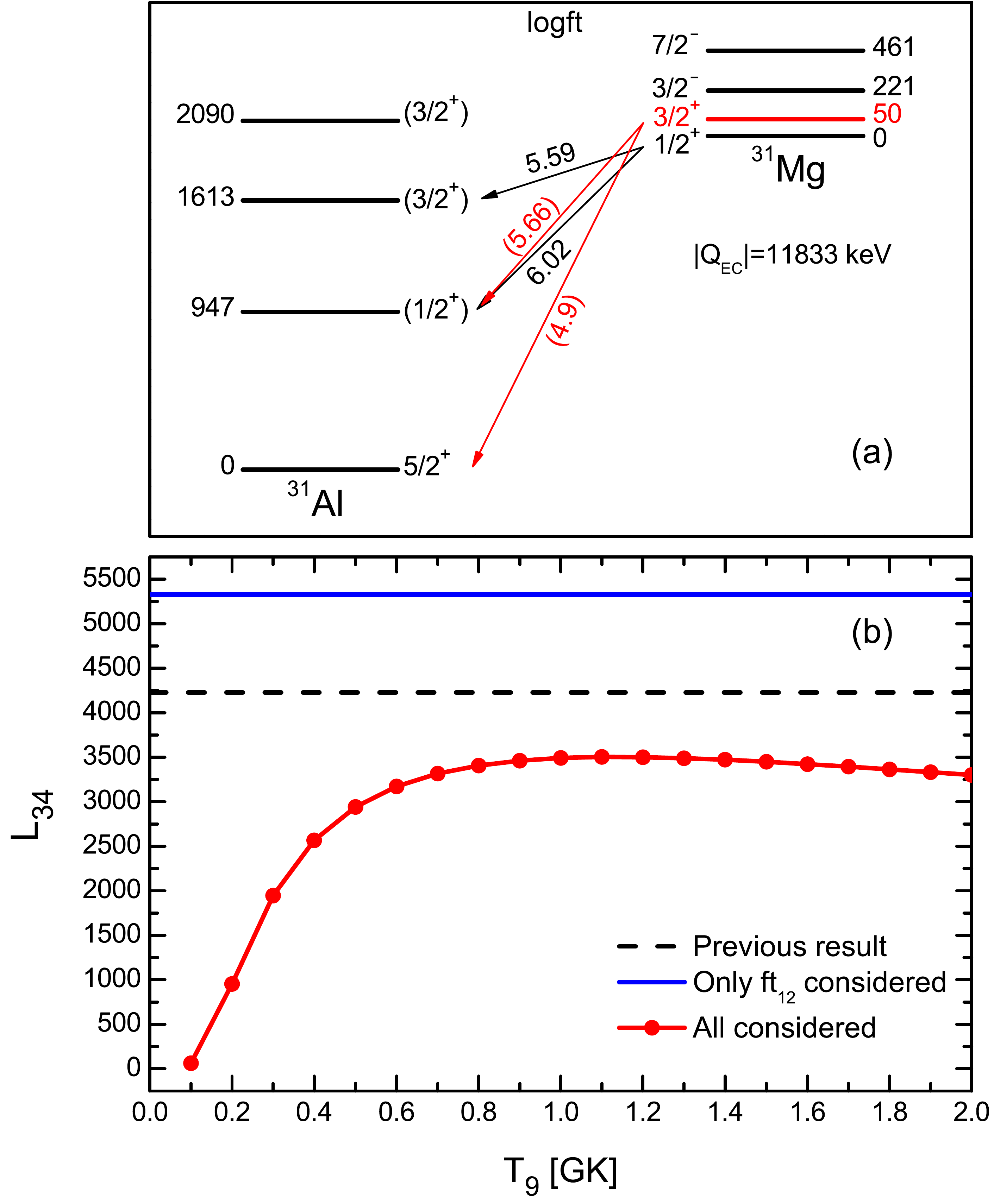}
  \caption{\label{fig:two} (Color online) (a). Schematic nuclear energy-level diagrams for the $^{31}$Al-$^{31}$Mg EC/$\beta^-$-decay Urca pair for neutron star crust. The energies (in keV) and $J^\pi$ for the ground states and lowest-lying states, as well as the available data and calculated results (in parenthesis) for log$ft$ of transitions in $\beta^-$-decays between different states are shown. (b). The corresponding $L_{34}$ calculated by Eqs. (\ref{previous_L_34}, \ref{got_L_34}) as functions of temperature $T_9$. Calculations with all transitions considered are compared with those with one transition as well as the previous result. See text for details. }
\end{center}
\end{figure}

With our modified expression, we first analyze the $^{31}$Al-$^{31}$Mg Urca pair as an example in the crust. This pair is expected to have a large mass fraction \cite{Meisel_2017_ApJ} and considerable neutrino luminosity ($\approx 10^{37}$ erg s$^{-1}$) \cite{Schatz_2014_Nature}, and it would therefore play an important role in crust cooling. Figure \ref{fig:two}(a) shows schematic level diagrams for low-lying states in $^{31}$Al and $^{31}$Mg. The excitation energies (keV), spin-parity assignments $J^\pi$, log$ft$ values, and EC $Q$-values $|Q_{\text{EC}}|$ are taken from evaluated data where available (ENSDF \cite{ENSDF} for energy, spin/parity, and log$ft$;  AME2016 \cite{AME2016} for atomic masses).  We supplement evaluated data with theoretical values (in parenthesis) calculated from the projected shell model (PSM) \cite{PSM-review, PSM-Sun, PSM-Sun2, Wang-2014-R, Wang_2016_PRC, Wang_2018_PRC, Tan_2020_PLB}; we adopt a quenching factor of $f_{\text{quench}}=0.85$ for the reduced probability of nuclear transitions \cite{Wang_2018_0vbb, Wang_2018_PRC, Gysbers_2019_Nat_Phys, Tan_2020_PLB}. 

Figure \ref{fig:two}(b) displays our calculated $L_{34}$ as a function of temperature based on the data in Fig. \ref{fig:two}(a). The results calculated from our Eq. (\ref{got_L_34}) are labeled as `All considered', which take nuclear excitations into account by considering all transitions between states of $^{31}$Al and $^{31}$Mg.  As in the literature, we use the approximation $ft_{\varepsilon \beta}^+ = ft_{\varepsilon \beta}^-$; we take log$ft^-=10$ for forbidden transitions.  We also show results calculated from Eq. (\ref{previous_L_34}) using $Q=Q_{\text{EC}}$ (difference of atomic masses, `Previous result') and $Q=Q_{12}$ (difference of nuclear energies, `Only $ft_{12}$ considered'), where only the transition between the ground state of $^{31}$Al ($\varepsilon = 1$) and the first excited state of $^{31}$Mg ($\beta = 2$) is considered (with log$ft^-_{12}=4.9$).  Fig. \ref{fig:two}(b) demonstrates that with only one transition,  $L_{34}$ increases moderately when $Q_{\text{EC}}$ is replaced by the more appropriate $Q_{12}$; this effect will be greater for smaller values of $Q_{\text{EC}}$ because the correction will be comparatively larger.

When nuclear excitations are considered, $L_{34}$ becomes strongly temperature-dependent as the state occupation probabilities follow the Boltzmann distribution. At temperatures below $T_9 = 0.1$, both $^{31}$Al and $^{31}$Mg stay mainly in their ground states, and transitions between them are forbidden; this results in a small $L_{34}$. With increasing temperature, $^{31}$Mg has increasing probability to be in its first excited state at $50$ keV.  The transition between this state and the $^{31}$Al ground state is allowed with a large matrix element (log$ft^-_{12}=4.9$), and as the occupation of this level increases, $L_{34}$ grows rapidly. At $T_9 \approx 1$, both the ground state and the first excited state of $^{31}$Mg have considerable occupations, although the temperature is not sufficient to significantly populate the second excited state at 221 keV.  Between $T_9 \approx 1$ and $T_9 \approx 1.5$, $L_{34}$ stays almost independent of temperature. Above $T_9 \approx 1.5$, the 221 keV state of $^{31}$Mg begins to be populated; transitions between this state and the low-lying states of $^{31}$Al are all forbidden, leading to a slight decrease in $L_{34}$.  At all temperatures, our calculations of $L_{34}$ are substantially lower than the previous values.

This analysis shows that low-lying nuclear excitations are crucial in the computation of $L_{34}$ and Urca neutrino luminosity. In the $^{31}$Al-$^{31}$Mg pair, correct treatment of nuclear excitations tends to reduce $L_{34}$ relative to previous results, and the reduction depends on the temperature of the environment. In the NS crust, at a typical temperature $T_9=0.51$ \cite{Schatz_2014_Nature}, the new $L_{34}$ for the $^{31}$Al-$^{31}$Mg pair is lower than the previous value by $\sim 40\%$.  This pair is expected to have a large mass fraction \cite{Meisel_2017_ApJ} and sizable neutrino luminosity ($\approx 10^{37}$ erg s$^{-1}$) \cite{Schatz_2014_Nature}, so the excited states will significantly modify understanding of Urca cooling in NS crusts.

\begin{figure}[htbp]
\begin{center}
  \includegraphics[width=0.45\textwidth]{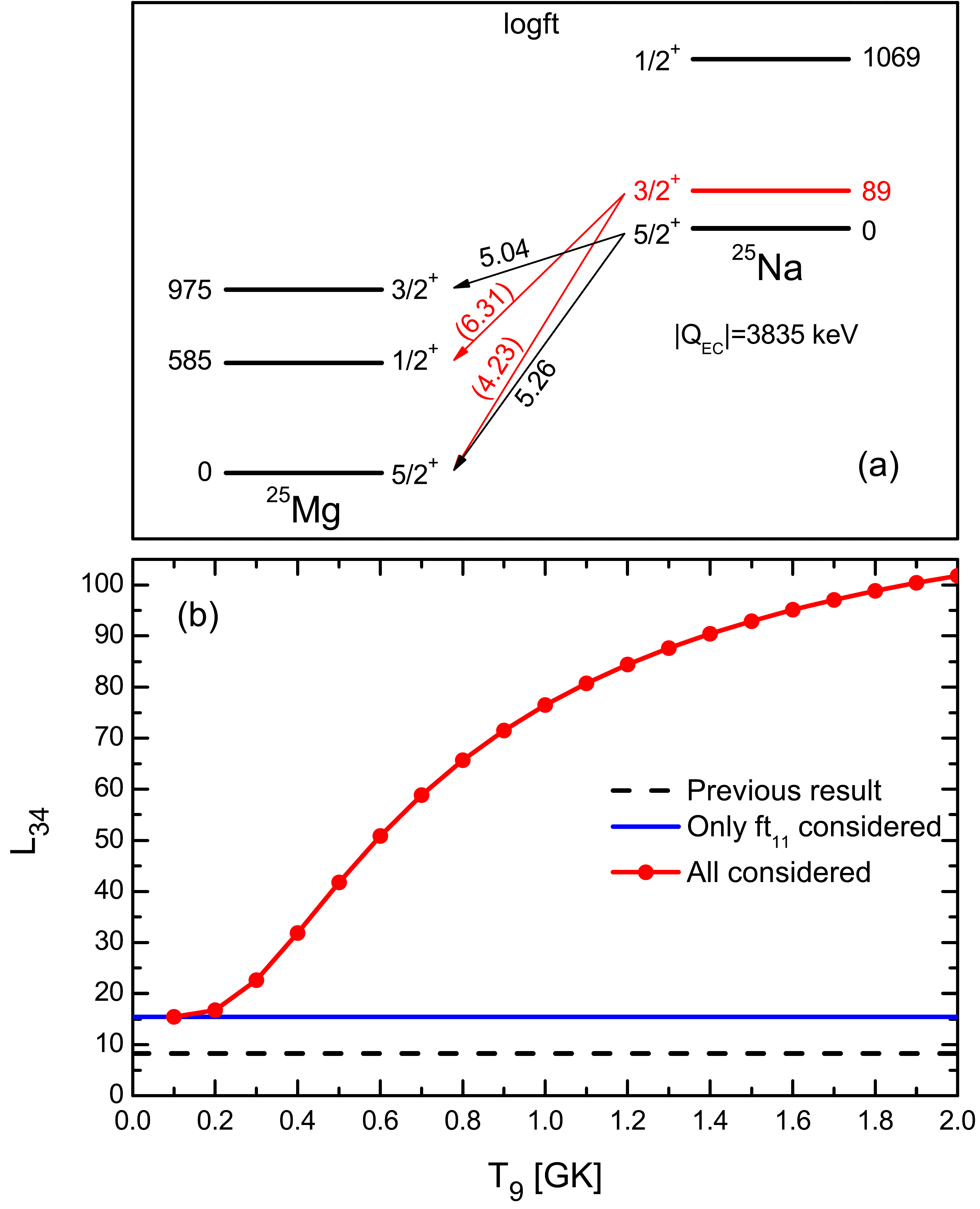}
  \caption{\label{fig:three} (Color online) The same as Fig. \ref{fig:two} but for the $^{25}$Mg-$^{25}$Na EC/$\beta^-$-decay Urca pair for neutron star ocean. }
\end{center}
\end{figure}

As an example in the NS ocean, we take the $^{25}$Mg-$^{25}$Na Urca pair, which is also expected to contribute sizable neutrino luminosities. In Fig. \ref{fig:three}(a) we show schematic nuclear level diagrams for $^{25}$Mg and $^{25}$Na. All data are taken from Ref. \cite{ENSDF}, except for the log$ft$ values for transitions from the first excited state of $^{25}$Na at 89 keV; we calculated those values (in parentheses) using the PSM.

Fig. \ref{fig:three}(b) shows the $L_{34}$ curves calculated from Eqs. (\ref{previous_L_34}, \ref{got_L_34}).  Using Eq. (\ref{previous_L_34}) and $Q_{\text{EC}}$, $L_{34}=8.25$. When $Q_{\text{EC}}$ is replaced by the more appropriate $Q_{11}$, $L_{34}$ nearly doubles to $15.43$ because of the comparatively small $|Q_{\text{EC}}|$ = 3.835 MeV. When transitions among the low-lying states are included using Eq. (\ref{got_L_34}), $L_{34}$ has a prominent temperature dependence: increasing temperature drives a large increase in $L_{34}$. This is because the $\beta$-decay parent nucleus $^{25}$Na has a rapidly-growing occupation of its first excited state at 89 keV. The transition rate between this and the ground state of $^{25}$Mg is predicted to be large, with log$ft^-_{12}=4.23$. At $T_9 \approx 1$, we predict $L_{34} \approx 70$, which is almost one order of magnitude greater than previous calculations \cite{Deibel_2016_ApJ}. This would make the $^{25}$Mg-$^{25}$Na pair the largest contributor to neutrino luminosity and Urca cooling in the NS ocean.

\begin{figure}[htbp]
\begin{center}
  \includegraphics[width=0.45\textwidth]{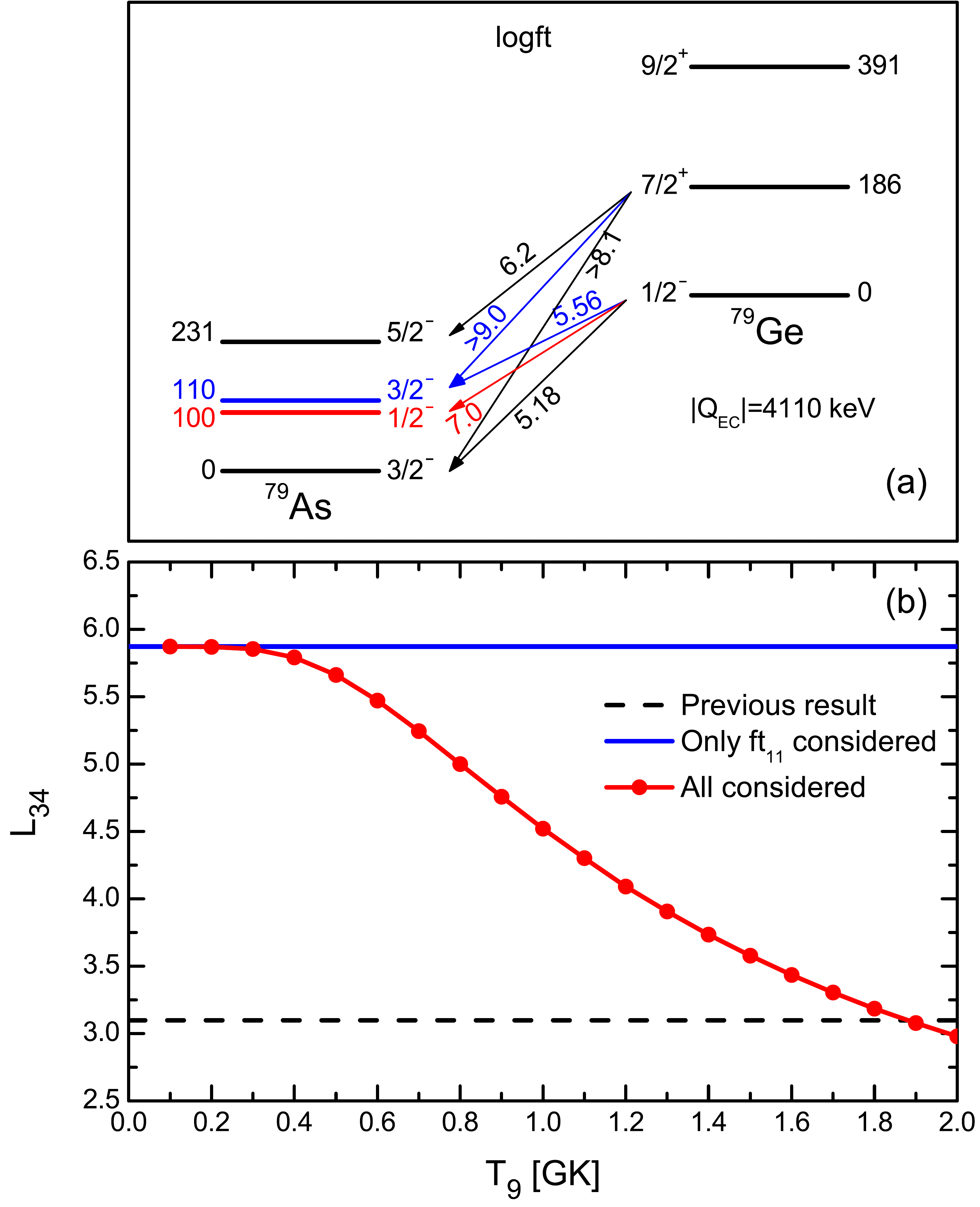}
  \caption{\label{fig:four} (Color online) The same as Fig. \ref{fig:two} but for the $^{79}$As-$^{79}$Ge EC/$\beta^-$-decay Urca pair for neutron star ocean. }
\end{center}
\end{figure}

Including more states does not always lead to an enhancement in $L_{34}$; the $^{79}$As-$^{79}$Ge Urca pair illustrates this point.  Fig. \ref{fig:four} shows our calculations of this pair; the nuclear properties are all from Ref. \cite{ENSDF}. In this case, the transition between the ground states is allowed and has the largest strength (smallest log$ft$) among low-lying transitions. Considering only this transition in Eq. (\ref{previous_L_34}) and using $Q_{EC}$, we find $L_{34} \approx 3.1$. If $Q_{EC}$ is replaced by $Q_{11}$, $L_{34}$ nearly doubles to $\approx 5.9$.  With the low-lying states included, $L_{34}$ again exhibits a sensitive temperature dependence.  However, in contrast with the other examples, the main trend of $L_{34}$ in this pair is to decrease as temperature increases; this is because the ground state occupation probabilities drop in favor of excited states with smaller transition strengths and hence lower rates.


To summarize, neutrino cooling is important in many astrophysical environments; strong neutrino emission carries away large amounts of energy.  Previous studies of the Urca process in NS crusts and oceans \cite{Schatz_2014_Nature, Deibel_2016_ApJ, Meisel_2017_ApJ} calculated neutrino luminosities from an expression that considers only one weak-interaction transition for each Urca pair. We found that this is a serious oversimplification because Urca pairs consist of odd-$A$ nuclei with low-lying states that can be thermally populated.

We derived an expression for the nuclear part of Urca neutrino luminosity ($L_{34}$) that explicitly includes excited states in both nuclei of an Urca pair.  Using our new expression, we studied the effects of nuclear excitations on the Urca process in NS crusts and oceans.  Our predicted neutrino luminosities may be substantially enhanced or suppressed relative to previous calculations; we found effects up to one order of magnitude.  The precise corrections depend sensitively on the detailed nuclear structure and weak-interaction transition strengths, as well as the temperature of the environment.

We expect that many medium-heavy neutron-rich nuclei which have sizable mass fractions \cite{Meisel_2017_ApJ} and low-lying states (due to variations in nuclear shape)---such as in the $A \approx 80$ region (see Fig. 2 of Ref. \cite{Schatz_2014_Nature})---will have substantial changes in their Urca cooling efficiency with the inclusion of excited states. This could amend the current understanding of Urca cooling in NS crusts and oceans.

A few additional comments are in order. First, the present work discussed only Urca pairs chosen from the candidate list in Ref. \cite{Schatz_2014_Nature, Deibel_2016_ApJ}; the consideration of excited states will likely alter conclusions about the most influential pairs.  Second, in the present work, we treated the isomer in $^{79}$Ge the same as a normal excited state.  Nuclear isomers can behave differently in thermal environments \cite{Sun_2005_Nat_Phys}, and they may fail to reach thermal equilibrium \cite{Banerjee_2018_PRC_isomer, Wendell_2020}; this warrants investigation into their roles in the Urca process. Third, although we specialized our analysis to NSs, our conclusions are valid generally for other astrophysical problems \cite{Janka_2000_NPA, Hix_2010_NPA}. Finally, our work may stimulate experiments to measure weak transition rates of excited states \cite{Ong_2020_PRL} and encourage more theoretical study of the Urca cooling mechanism in many astrophysical environments \cite{Schwab_2017_ApJ, XPZ_2017_PRC}.

\begin{acknowledgments}
The authors thank George Fuller for valuable discussions and communications. This work is supported by the National Natural Science Foundation of China (Grant Nos. 11905175, 11875225, and U1932206), by the National Key Program for S$\&$T Research and Development (Grant No. 2016YFA0400501), by the Venture $\&$ Innovation Support Program for Chongqing Overseas Returnees (with Grant No. cx2019056), and by the Fundamental Research Funds for the Central Universities (with Grant No. SWU019013).
\end{acknowledgments}


\providecommand{\noopsort}[1]{}\providecommand{\singleletter}[1]{#1}%

\end{document}